\documentclass{gtmon_a}
\pdfoutput=1

\proceedingstitle{The interaction of finite-type and Gromov--Witten
invariants (BIRS 2003)}
\conferencestart{15 November 2003}
\conferenceend{20 November 2003}
\conferencename{The interaction of finite-type and Gromov--Witten
invariants}
\conferencelocation{Banff International Research Station, Banff, Alberta,
Canada}

\editor{David Auckly}
\givenname{David}
\surname{Auckly}

\editor{Jim Bryan}
\givenname{Jim}
\surname{Bryan}

\title{Open string instantons and relative stable morphisms}

\author{Jun Li}
\givenname{Jun}
\surname{Li}
\address{Department of Mathematics\\Stanford University\\\newline
Stanford\\
California 94305\\USA}
\email{jli@math.stanford.edu}

\author{Yun S Song}
\givenname{Yun S}
\surname{Song}
\address{Department of Physics, and SLAC\\
Stanford University\\\newline
Stanford\\
California 94305\\USA}
\email{yss@leland.stanford.edu}

\volumenumber{8}
\issuenumber{}
\publicationyear{2006}
\papernumber{2}
\startpage{49}
\endpage{72}

\doi{}
\MR{}
\Zbl{}
\keyword{open string theory}
\keyword{Gromov-Witten invariants}
\keyword{large N duality}
\subject{primary}{msc2000}{14N35}
\subject{secondary}{msc2000}{14D21}

\received{26 April 2001}
\revised{}
\accepted{}
\published{22 April 2006 (in these proceedings)}
\publishedonline{22 April 2006 (in these proceedings)}
\proposed{}
\seconded{}
\corresponding{}
\editor{}
\version{}

\arxivreference{arXiv:hep-th/0103100}

\dedicatory{Reproduced by kind permission of International Press
from:\newline 
{\rm Advances in Theoretical and Mathematical Physics, Volume 5 (2002) pages 69--91}}

\AtBeginDocument{\let\bar\wbar\let\tilde\wtilde\let\hat\what\let\sref\fullref
\let\eqr\ref}
\allowdisplaybreaks

\numberwithin{equation}{section}

\def\inbar{\,\vrule height1.5ex width.4pt depth0pt}

\def\IC{{\relax\hbox{$\inbar\kern-.3em{\rm C}$}}}

\def\IP{{\mathbf P}}
\def\sIP{\mbox{\bf \scriptsize P}}
\def\calo{{\cal O}}

\def\mtiny#1{\mbox{\tiny {$#1$}}}
\def\mscript#1{\mbox{\scriptsize {$#1$}}}
\def\wt{\tilde{t}}

\DeclareMathOperator{\Aut}{Aut}

\DeclareMathOperator{\Hom}{Hom}
\DeclareMathOperator{\Ext}{Ext}

\def\cO{{\mathcal O}}

\def\cU{{\mathcal U}}

\def\bC{{\mathbf C}}

\def\bM{{\mathbf M}}

\def\bd{{\mathbf d}}

\def\bone{{\boldsymbol 1}}

\def\ZZ{{\mathbb Z}}

\def\RR{{\mathbb R}}
\def\CC{{\mathbb C}}

\newtheorem{prop}{Proposition}[section]

\def\and{\quad{\rm and}\quad}

\def\bd{{\mathbf d}}

\def\bone{{\mathbf 1}}

\def\dual{^{\vee}}

\def\defeq{:=}

\def\eps{\epsilon}

\let\lra=\longrightarrow

\let\mh\co

\def\mgdd{\bM_{g,\mu}\urel}

\def\onec{\bone_\CC}
\def\pri{^{\prime}}

\def\Po{{\IP^1}}
\def\Pm{\Po[m]}

\def\qi{q_\infty}
\def\qz{q_0}

\def\sub{\subset}
\def\sta{^{\ast}}

\def\upmo{^{-1}}

\def\urel{^{\text{rel}}}

\def\vir{^{{\rm vir}}}

\newcommand{\Om}{\Omega}
\newcommand{\om}{\omega}
\newcommand{\ocal}{{\cal O}}
\newcommand{\ra}{\rightarrow}
\newcommand{\iof}{{i_1}_*}
\newcommand{\itf}{{i_2}_*}
\newcommand{\mH}{\mbox{H}}
\newcommand{\pdd}[1]{{\partial \over \partial #1}}

\def\yboxit#1#2{\vbox{\hrule height #1 \hbox{\vrule width #1
\vbox{#2}\vrule width #1 }\hrule height #1 }}
\def\fillbox#1{\hbox to #1{\vbox to #1{\vfil}\hfil}}
\def\ybox{{\lower 1.3pt \yboxit{0.4pt}{\fillbox{8pt}}\hskip-0.2pt}}
\begin{document}

\begin{asciiabstract}
We show how topological open string theory amplitudes can
be computed by using relative stable morphisms in the algebraic category.
We achieve our  goal by explicitly working through an example which has
been previously considered by Ooguri and Vafa from the point
of view of physics.  By using the method of virtual localization, we
successfully reproduce their results for multiple covers of a holomorphic
disc, whose boundary lies in a Lagrangian submanifold of a Calabi-Yau
3-fold, by Riemann surfaces with arbitrary genera and number of
boundary components.  In particular we show that 
in the case we consider there are no open string instantons with more
than one boundary component ending on the Lagrangian submanifold.
\end{asciiabstract}

\begin{abstract}
We show how topological open string theory amplitudes can
be computed by using relative stable morphisms in the algebraic category.
We achieve our  goal by explicitly working through an example which has
been previously considered by Ooguri and Vafa from the point
of view of physics.  By using the method of virtual localization, we
successfully reproduce their results for multiple covers of a holomorphic
disc, whose boundary lies in a Lagrangian submanifold of a Calabi--Yau
3--fold, by Riemann surfaces with arbitrary genera and number of
boundary components.  In particular we show that 
in the case we consider there are no open string instantons with more
than one boundary component ending on the Lagrangian submanifold.
\end{abstract}

\maketitle


\renewcommand{\thefootnote}{\fnsymbol{footnote}}

\renewcommand{\thefootnote}{\arabic{footnote}}
\setcounter{footnote}{0} \setcounter{section}{-1}


\section{Introduction}
The astonishing link between intersection theories on moduli spaces
and topological {\em closed} string theories has by now taken a
well-established form, a progress for which E Witten first plowed the
ground in his seminal paper \cite{Witten1}.
As a consequence, there now exist rigorous
mathematical theories of Gromov--Witten invariants, which
naturally arise in the aforementioned link.  In the symplectic
category, Gromov--Witten invariants were first constructed for
semi-positive symplectic manifolds by Y Ruan and G Tian \cite{RT}.
To define the invariants in the algebraic
category, J Li and G Tian
constructed the virtual fundamental class of the moduli space of
stable maps by endowing the moduli space with an extra structure called
a perfect tangent-obstruction complex \cite{LT2}.\footnote{Alternative
constructions were also made by Y Ruan \cite{Ru} and by B Siebert \cite{Si}.}
Furthermore, Gromov--Witten theory
was later extended to general symplectic manifolds by Fukaya and Ono \cite{FO},
and by Li and Tian \cite{LT1}. 
In contrast to such an impressive list of advances just described, no
clear link currently exists between topological {\em open}  string
theories and intersection theories on moduli spaces.  One of the most
formidable obstacles that stand in the way to progress is that
it is not yet known how to construct well-defined moduli spaces of
maps between manifolds with boundaries.
The main goal of this paper is to contribute to narrowing the existing
gap between topological open string theory and
 Gromov--Witten theory.  In so doing we hope that our
work will serve as a stepping-stone that will take us a bit closer to
answering how relative stable morphisms can be used to study
topological open string theory.

In order to demonstrate the proposed link between topological open
string theory and Gromov--Witten theory, we will focus on an
explicit example throughout the paper.  The same example was
also considered by string theorists Ooguri and
Vafa in \cite{OV}, where they used  results from
Chern--Simons theory and M--theory to give two independent derivations
of open string instanton
amplitudes.  A more detailed description of the problem will be
presented later in the paper.  We just mention here that, by using our
mathematical approach, we have successfully reproduced their answers
for multiple covers of a holomorphic disc by Riemann surfaces of
arbitrary genera and number of holes.  In fact we show that there are
no open string instantons with more than one hole, a result which was
anticipated in \cite{OV} from their physical arguments.

The invariants we compute are a generalization of {\em absolute}
Gromov--Witten invariants that should be more familiar to string
theorists.  Our case involves relative stable maps which intersect a
specified complex-codimension-two submanifold of the target space in a
finite set of points with multiplicity.  It will become clear later in
the paper that the theory of relative stable maps is tailor-made for
studying topological open string theory.  The construction of relative
stable maps was first developed in the symplectic category (Li--Ruan
\cite{LR}, Ionel--Parker \cite{IP1,IP2}).  Recently in \cite{Li1,Li2}
the first author of the present paper has given an algebro-geometric
definition of the moduli space of relative stable morphisms and has
constructed relative Gromov--Witten invariants in the algebraic
category.  The foundation of our work will be based on those papers.

The organization of this paper is as follows:
In \sref{s:description} we give a brief description of the multiple
cover problem that arose in \cite{OV} and state what we wish to
reproduce using relative stable morphisms.
In \sref{s:modulispace} we define the moduli space of relative stable
morphisms and describe how multiple covers of a holomorphic disc can
be viewed as a problem regarding relative stable morphisms.
We investigate the obstruction theory of the moduli space in
\sref{s:obstruction}. 
In \sref{s:localization}
we study the localization of the virtual fundamental class
and compute the equivariant Euler class of the
virtual normal bundle to the fixed locus.
In \sref{s:RelGWh=1} we evaluate the relevant invariants for the case
where the source Riemann surface has only one boundary component. 
The cases with more than one boundary component are subsequently discussed in
\sref{s:RelGWh>1}.
We conclude in \sref{s:Conc} with some comments.

{\bf Note}\qua While this manuscript was in its final stage of
preparation, Katz and Liu announced their results
\cite{KL} which deal with the same subject matter as our paper.


\section{A brief description of the problem}
\label{s:description}

The notion of duality has been one of the most important common
threads that run through modern physics.  A duality draws intricate
connections between two seemingly unrelated theories and often allows
one to learn about one theory from studying the other.  A very
intriguing duality correspondence has been proposed by Gopakumar and
Vafain \cite{GV}, where the authors provide several supporting
arguments for a duality between the large--$N$ expansion of $SU(N)$
Chern--Simons theory on $S^3$ and a topological {\em closed} string
theory on the total space of the vector bundle
$\calo_{\sIP^1}(-1)\oplus \calo_{\sIP^1} (-1)$ over
$\IP^1$.\footnote{See \cite{GV} and references therein for a more
precise account of the proposal.}  The equivalence was established in
\cite{GV} at the level of partition functions.  We know from Witten's
work in \cite{Witten2}, however, that there are Wilson loop
observables in Chern--Simons theory which correspond to knot
invariants.  The question then is, ``What do those invariants that
arise in Chern--Simons theory correspond to on the topological string
theory side?''

The first explicit answer to the above question was  given by Ooguri
and Vafa in \cite{OV}.  In the case of a simple knot on
$S^3$, by following through the proposed duality in close detail, they
showed that the corresponding quantities on the topological string
theory  side are open string instanton amplitudes.  More precisely, in the
particular example they consider, the open string instantons map to
either the upper or the lower hemisphere of the base
$\IP^1$.\footnote{We clarify that the geometric set up in the present
case is no longer that described above.  There is a unique Lagrangian
3--cycle ${\cal C}_K$ in $T^*S^3$ which intersects $S^3$ along a given knot
$K$ in $S^3$.  Associated to such a 3--cycle ${\cal C}_K$ in $T^*S^3$
there is a Lagrangian 3--cycle $\tilde{{\cal C}}_K$ in the local Calabi--Yau
three-fold $X$ of the topological string theory side.  For the 
simple knot $S$ considered by Ooguri and Vafa, the latter 3--cycle
$\tilde{{\cal C}}_S$ intersects the base $\IP^1$ of $X$ along its
equator.  It is  the presence of this 3--cycle that allows for
the existence of holomorphic maps from Riemann surfaces with
boundaries to either the upper or the lower hemisphere.  See
\cite{OV} for a more detailed discussion.}

According to \cite{OV},
the generating function for topological open string amplitudes
is
\begin{equation}
F(t,V) =
\sum_{g=0}^\infty\sum_{h=0}^\infty\sum_{d_1,\ldots,d_h}^\infty
\lambda^{2g-2+h} F_{g;d_1,\ldots,d_h}(t)
\prod_{i=1}^{h}\mbox{tr}V^{d_i}, \label{eq:GenF}
\end{equation}
where $t$  is the K\"ahler modulus of $\IP^1$; $V$ is a path-ordered
exponential of the gauge connection along the equator and
$\mbox{tr}V^{d_i}$ arises from  the $i^{\rm th}$
boundary component which winds around the equator $|d_i|$--times with
orientation, which determines the sign of $d_i$;  $\lambda$ is the
string coupling constant; and  $F_{g;d_1,\ldots,d_h}$ is the
topological open string amplitude on a genus--$g$ Riemann surface with
$h$ boundary components.  Furthermore, by utilizing the aforementioned duality
with Chern--Simons theory, Ooguri and Vafa concluded that 
\begin{equation}
F(t,V) = i \sum_{d=1}^\infty {\mbox{tr}V^{d} + \mbox{tr}V^{-d}\over
2d\, \sin({d\lambda/2})} e^{-dt/2}, \label{eq:FOV}
\end{equation}
which they confirmed by using an alternative approach in the
M--theory limit of type IIA string theory.\footnote{We refer the reader
to the original reference \cite{OV} for further description of this approach.}
By comparing \eqr{eq:GenF} and \eqr{eq:FOV}, one immediately sees that
there are no open string instantons with more than one boundary
component ending on the equator;
that is, $F_{g;d_1,\ldots,d_h}=0$ for $h>1$.  To extract the
topological open string amplitude on a genus--$g$ Riemann surface with
one boundary component $(h=1)$, we need to expand \eqr{eq:FOV} in powers of
$\lambda$.  After some algebraic manipulation, we see that 
\begin{eqnarray*}
F(t,V) 
&=& i \sum_{d=1}^\infty \left( {1\over d^2} \lambda^{-1} +
\sum_{g=1}^\infty  d^{\,2g-2} 
{2^{2g-1} -1 \over 
2^{2g-1}} {|B_{2g}|\over (2g)!} \lambda^{2g-1} \right)\\
&&\hspace{2.5in} e^{-dt/2} \left(\mbox{tr}V^{d} +
\mbox{tr}V^{-d} \right),
\end{eqnarray*}
where $B_{2g}$ are the Bernoulli numbers defined by
\[
\sum_{n=0}^\infty B_n {x^n\over n!} \ = \ {x\over e^x -1}.
\]
Hence, topological open string amplitudes, which correspond to
multiple covers of either the upper or the lower hemisphere
inside the local Calabi--Yau three-fold described above, are
\begin{equation}
-i F_{g;d_1,\ldots,d_h}(0) = \left\{ \begin{array}{ll} 
     {d^{-2}} ,  & g=0,h=1, |d_1|=d > 0, \\ \displaystyle
     d^{\,2g-2} \left({2^{2g-1} -1 \over 2^{2g-1}} {|B_{2g}|\over (2g)!}\right)
                ,     & g>0, h=1, |d_1|=d>0,\\
      0 , & \mbox{otherwise}.
     \end{array} \right. \label{eq:Fgh}
\end{equation}
In the remainder of this paper, we will work towards reproducing these
results using relative stable morphisms.

\section{Relative stable maps and Lagrangian submanifolds}
\label{s:modulispace}

In this section, we will formulate the problem in terms of the
moduli space of stable holomorphic maps. 
Throughout this paper, we fix two points $q_0$ and
$q_\infty\in\Po$, and choose a
homogeneous coordinate $[w_1,w_2]$ of $\Po$ so that $\qz$ (resp.
$\qi$) is the point $[0,1]$ (resp. $[1,0]$). We will use $w$ to
denote the standard coordinate of $\bC=\Po-\qi$ so that $w$ is
related to the homogeneous coordinate $[w_0,w_1]$ via $w=w_0/w_1$.
Hence $\qz$ and $\qi$ become $0$ and $\infty$, respectively.
We let $W$ be the total space of
the vector bundle $\onec\oplus\onec$ over $\bC=\Po-q_\infty$. (In this
paper, we will use $\bone_X$ to denote the trivial holomorphic
line bundle over $X$.)
For any $r\in\RR^+$ we denote by $D_r\sub\bC$ the closed disk $|w|\leq r^d$
and denote the boundary of $D_r$ by $\partial D_r$.  Furthermore, we let
$W_r=\pi\upmo(D_r)$, where $\pi\mh W\to\bC$ is the projection. In
the present section, we will consider the space of holomorphic maps
from Riemann surfaces to $W_r$ whose boundary lies in a Lagrangian
submanifold in $\partial W_r$.

We now describe this Lagrangian submanifold, following Ooguri and Vafa
\cite{OV}. 
We let $e_1$ and $e_2$ be the constant $1$ section of the first and
the second factor of 
$\onec\oplus\onec$. Then any vector $\xi\in \onec\oplus\onec$ over
$s\in \bC$ is expressed uniquely as $\xi=(s,u e_1+v e_2)$. In case $w$
is the coordinate of $s\in\bC$, we say $(w,u,v)$ 
is the coordinate of $\xi\in W$. The Lagrangian submanifold introduced
by Ooguri and Vafa in \cite{OV} is\footnote{The authors of \cite{OV}
considered the Lagrangian submanifold $u_-=w\bar v_-$, where $u_-$ and
$v_-$ are given as sections of $\cO_{\Po}(-1)$ using a frame over
$\Po-\qz$. The transition between $u$ and $u_-$ is $u_-=w u$. 
Under this condition, our equation is equivalent to that of Ooguri and Vafa.}
\begin{equation}
\label{10.1}
B_r=\{(w,u,v)\mid |w|=r^d,\ u=\bar w\bar v\}.
\end{equation}
As a convention in this paper, we will use $t=e^{i\theta}$ to denote a
general element in 
the group $S^1$. We fix an $S^1$--action on $\bC\sub\Po$ via
\begin{equation}
\label{10.2}
[w,1]^t=[tw,1].
\end{equation}
Then $w^t=t\upmo w$, where $w^t$ is the push-forward of the function
$w$ under the group action, 
which is the pull-back of $w$ under the inverse of \ref{10.2}.
We next choose an $S^1$--linearization of $W\to\bC$ so that it leaves
$B_r$ invariant. We choose
\begin{equation}
\label{11.1}
e_1^t=te_1\quad\text{and}\quad e_2^t=e_2.
\end{equation}
Then $(s,ue_1+ve_2)^t=(s^t,u e_1^t+ve_2^t)$, or equivalently
$$(w,u,v)^t=(t\upmo w,tu, v).
$$
Hence $(u-\bar w\bar v)^t=tu-t\bar w\bar v$ and thus $B_r$ is $S^1$--invariant.

We now define the moduli space $\mgdd(W_r,B_r)$ of relative
holomorphic maps from genus--$g$ Riemann surfaces 
with boundaries to $(W_r,B_r)$, with prescribed winding numbers along
the boundaries 
of the domain Riemann surfaces.
We let $g\geq 0$ be an integer, and  $\mu=(d_1,\cdots,d_n)$ an
$n$--tuple of positive integers. In this paper, 
we will call connected, holomorphic nodal curves with ordered smooth
boundaries prestable nodal holomorphic curves with ordered boundaries.
When $\Sigma$ is such a curve we will use $\partial_k \Sigma$ to
denote the $k^{\rm th}$ boundary component of $\Sigma$. Naturally, we give
each boundary  $\partial_k\Sigma$ its induced orientation. 
Now let $\Sigma$ be a holomorphic nodal curve with $n$ ordered
boundary components and let $f\mh (\Sigma,\partial\Sigma)\to (W,B_r)$
be a holomorphic map. We say $f$ has winding number $d_k$ along the $k^{\rm th}$
boundary component if the degree of
$$\pi\circ f|_{\partial_k\Sigma}: \partial_k\Sigma \to \partial D_r
$$
has degree $d_k$.
We say $f$ is a relative holomorphic map (with $\mu=(d_1,\cdots,d_n)$
implicitly understood) 
if, for all $k$, it has winding number $d_k$
along its $k^{\rm th}$ boundary component. A relative holomorphic map
is said to be stable if there is no irreducible component of $\Sigma$
that is isomorphic to $\Po$ that 
contains only one nodal point and that is mapped to a single point in $W_r$.
As usual, two such relative stable maps $(f_1,\Sigma_1)$ and
$(f_2,\Sigma_2)$ are equivalent if there is an isomorphism $\rho\mh
\Sigma_1\to\Sigma_2$ that preserves the order of the boundary
components so that $f_1=\rho\circ f_2$.
We define $\mgdd(W,B_r)$ to be the moduli space of all relative stable
maps from genus--$g$ curves
to $(W_r,B_r)$, modulo equivalence.
Similarly, we define $\mgdd(D_r)$ to be the moduli space of all
relative stable maps from genus--$g$ curves
with ordered boundaries
to $(D_r, \partial D_r)$, modulo equivalence relation. Since $(D_r,
\partial D_t)\sub (W_r, B_r)$, $\mgdd(D_r)$ 
is a subspace of $\mgdd(W_r,B_r)$.

We now introduce another moduli space, the moduli space of relative
stable morphisms of genus--$g$ to $(\Po,\qi)$, with prescribed
ramification orders. Let $g$ and $\mu$ be as above. An {\sl ordinary}
relative stable morphism of genus--$g$ and of ramification order $\mu$
consists of a connected $n$--pointed nodal algebraic curve
$(C,,x_1,\cdots,x_n)$ and a morphism $f\mh C\to \Po$, so that  
$$f\upmo(\qi)=d_1x_1+\cdots+d_nx_n
$$
as a divisor. We say $(f,C)$ (with $\mu$ understood
implicitly) is a relative stable morphism if $f$ is stable as a
morphism from $C$ to $\Po$. 
We let $\mgdd(\Po)_0$ be the moduli space of {\sl ordinary}
relative stable morphisms, with ramification order $\mu$, from
genus--$g$ algebraic curves to $\Po$. It is
a Deligne--Mumford stack. In \cite{Li1}, the first author of the
present paper has constructed the moduli space 
$\mgdd(\Po)$ of relative
stable morphisms, with ramification order $\mu$, 
 from genus--$g$ algebraic curves to $\Po$.
The space $\mgdd(\Po)$ is a proper Deligne--Mumford
stack. It admits a natural perfect obstruction theory and hence admits
a virtual moduli cycle. 
The moduli stack $\mgdd(\Po)$ contains $\mgdd(\Po)_0$ as its open substack.
The new ingredient of
this moduli space is that a relative stable morphism $f\in\mgdd(\Po)$
consists of an $n$--pointed 
genus--$g$ algebraic curves $(C,x_1,\cdots,x_n)$ and a morphism $f\mh
C\to \Pm$, so that 
$$f\upmo(\qi)=
d_1x_1+\cdots+d_nx_n.
$$
Here $\Pm$ is the nodal curve that has $m$ ordered irreducible
components, each isomorphic 
to $\Po$, so that the $j^{\rm th}$ component intersects with the $(j+1)^{\rm th}$
component of $\Po[m]$ at 
exactly one point.
$\Po[m]$ also contains a distinguished divisor $\qi$ in the first
component of $\Po[m]$, and it
comes with a projection $\varphi\mh\Po[m]\to\Po$ that  maps
$\qi\in\Po[m]$ to $\qi\in\Po$.  Moreover, 
the restriction of $\varphi$ to the last component is an isomorphism
to $\Po$, and $\varphi$ contracts all other components of $\Po[m]$.
For any morphism $f\mh C\to \Pm$ we define $\Aut(f)\urel$ to be the set of
pairs $(a,b)$, where $a\mh C\to C$ are automorphisms of $(C,x_1,\cdots,x_n)$
and $b\mh \Pm\to\Pm$ are automorphisms\footnote{By this we mean that $b$ is
an isomorphism of 
$\Pm$ that preserves $\qi\in\Po[m]$ and the projection
$\varphi\mh\Po[m]\to\Po$.} 
so that $f\circ a=b\circ f$. We say $f$ is stable if $\Aut(f)\urel$ is finite.
From this description, we see that $\mgdd(\Po)_0\sub\mgdd(\Po)$ is the
open subset 
consisting of relative stable morphisms $f$ whose codomain is $\Po[1]=\Po$.


We now investigate the moduli space $\mgdd(W_r,B_r)$ of relative
stable holomorphic maps. 
Let $(f,\Sigma)\in \mgdd(W_r,B_r)$ be any
relative stable map. Composed with $\pi\mh W_r\to B_r$, we obtain a new map
$$\tilde f=\pi\circ f\co  \Sigma\to D_r.
$$
Since fibers of $W_r\to D_r$ are vector spaces, $(\tilde f,\Sigma)$ is
also relative stable and 
hence is in $\mgdd(D_r)$. This defines a map
$$\mgdd(W_r,B_r)\lra \mgdd(D_r).
$$
We now show that this is an isomorphism.
Given $f\in\mgdd(W_r,B_r)$ and its induced $\tilde f\in\mgdd(D_r)$,
the original $f$ is given by a holomorphic section
\begin{equation}
\label{10.4}
s\in \Gamma(\Sigma, \tilde f\sta L_1\oplus \tilde f\sta L_2).
\end{equation}
We now show that $s=0$. We let $\Sigma_2$ be the nodal curve $\Sigma$
with the conjugate 
holomorphic structure. Namely, $\Sigma_2$ is diffeomorphic to
$\Sigma$, its conformal structure 
on each irreducible components is identical to that of $\Sigma$, and
its orientation is the 
opposite of that of $\Sigma$.
We let $\Sigma_1$ be the identical copy of $\Sigma$.
We then let $\tilde \Sigma$ be the
nodal holomorphic curve derived by identifying the boundary of
$\Sigma_1$ with the boundary of 
$\Sigma_2$ using the identity map between their boundaries.
By the reflection principle, the complex structures of $\Sigma_1$ and
$\Sigma_2\sub\tilde\Sigma$ 
extend to a holomorphic structure on $\tilde\Sigma$.
The new curve $\tilde\Sigma$ is a nodal curve without boundary.
We next introduce a line bundle $\tilde L$ on $\tilde\Sigma$. We pick
a sufficiently small open 
neighborhood $V_i\sub\tilde\Sigma$ of $\Sigma_i\sub\tilde\Sigma$. We
let $\tilde L|_{V_i}\cong \cO_{V_i}$ and let $\zeta_i$ be the section
$1$ in $\cO_{V_i}$. 
The identification of $\tilde L|_{V_1}$ with $\tilde L|_{V_2}$ over
$V_1\cap V_2$ 
is given by
$$\zeta_1=f\sta(w)r^{-2d}\zeta_2.
$$
Now let $s$ in \ref{10.4} be given by $s(z)=u(z)e_1+v(z) e_2$.
Then $u(z)\zeta_1$ is naturally a holomorphic section of
$\tilde L|_{\Sigma_1}$ and $\bar v(z)\zeta_2$ also is a holomorphic section
of $\tilde L|_{\Sigma_2}$.
The condition that $f(z)\in B_r$ for $z\in\partial\Sigma$
implies that $u(z)=\overline{f(z)}\overline{v(z)}$ $z\in\partial\Sigma$.
Since $|f(z)|=r^d$, we have 
$$u(z)\zeta_1=\overline{f(z})\overline{v(z})f(z)r^{-2d}\zeta_2=\overline{v(z})\zeta_2.   
$$
This implies that $u\,\zeta_1$ and $\bar{v}\,\zeta_2$ patch together to form a
continuous section of $\tilde L$.
Since $u\,\zeta_1$ and $\bar v\,\zeta_2$ are holomorphic over
$\Sigma_1$ and $\Sigma_2$, respectively, 
$\tilde s$ is holomorphic. Finally, it is straightforward to check that
$\deg \tilde L=-\sum d_k<0$. Hence $\tilde s\equiv 0$, which shows
that $f\equiv \tilde f$. 
This proves that $\mgdd(W_t,B_r)=\mgdd(D_r)$
as sets. However, since the above construction works for analytic
families of relative stable 
maps in $\mgdd(W_r,B_r)$, this argument also shows that the two moduli spaces
$\mgdd(W_r,B_r)$ and $\mgdd(D_r)$ are isomorphic as analytic schemes (stacks).


We now show that $\mgdd(D_r)$ is naturally an open subset of $\mgdd(\Po)_0$.
Let $(f,\Sigma)$ be a relative holomorphic map
in $\mgdd(D_r)$. We let $D_r^{\epsilon}$ be $\{ (r-\eps)^d<|w|\leq
r^d\}$, which is a neighborhood of $\partial D_r\sub D_r$. Since
$f(\Sigma)\sub D_r$, $f(\partial \Sigma)\sub \partial D_r$ 
and $d_k>0$, the derivative of
$f$ along $\partial \Sigma$ is nowhere vanishing. Note that $f$ is
analytic up to the boundary by the reflection principle. Hence there
is a positive $\epsilon >0$ so that $f\upmo(D_r^{\epsilon})\to 
D_r^{\epsilon}$ is a covering space. Furthermore, if we let $U_k$ be
the connected component of 
$f\upmo(D_r^{\epsilon})$ containing $\partial_k\Sigma$,
then $U_k\to D_r^\epsilon$ is a $d_k$--fold covering.
Hence, there is a biholomorphic map
$$\rho_k\co 
\{(r-\epsilon)\upmo>|z_k|> r\upmo\}\lra U_k
$$
so that $f\circ \rho_k$ is given by $w=z_k^{-d_k}$. We then attach the
disk $\{|z_k|<(r-\epsilon)\upmo\}$ 
to $\Sigma$ using the isomorphism $\rho_k$. We let $C$ be the complete
nodal curve resulting from attaching $n$ disks to the $n$ holes,
following the procedure described. 
Then the map $f\mh \Sigma\to D_r$ extends to a holomorphic map $C\to
\Po$, which we denote by $\bar f$, 
so that $\bar f$ restricting to the disk $\{|z_k|<(r-\epsilon)\upmo\}$
is given by $w=z_k^{-d_k}$. Thus $(\bar f,C)$ is a relative stable
morphism in $\mgdd(\Po)$. 
It is straightforward to check that any two such extensions (to $(f,C)$) are
equivalent. This way we obtain a map
\begin{equation}
\label{10.3}
\mgdd(D_r)\lra\mgdd(\Po).
\end{equation}
This map is obviously injective and the image is an open subset in
$\mgdd(\Po)$. On the other hand, this construction works for any
analytic family of relative stable maps in $\mgdd(D_r)$. This shows
that \ref{10.3} is an open embedding. 
In the following, we will view $\mgdd(D_r)$ as an
open subset of $\mgdd(\Po)$. It is straightforward to check that, for
$r>r\pri$, $\mgdd(D_r)\sub \mgdd(\Po)$ is contained in
$\mgdd(D_{r\pri})\sub\mgdd(\Po)$. 
When we take the union of all these open subsets, we obtain
$$\bigcup_{r>0}\mgdd(D_r)=\mgdd(\Po)_0.
$$


\section{Obstruction theory of $\mgdd(W_r,B_r)$}
\label{s:obstruction}

We now investigate the obstruction theory of $\mgdd(W_r)$. First,
since $\mgdd(W_r)$ is identical to $\mgdd(B_r)$ as analytic schemes
(stacks), the space of the first order deformations of
$f\in\mgdd(W_r,B_r)$ is naturally isomorphic to the space of the first
order deformations of $f$ as an elements in $\mgdd(D_r)$.  As to the
obstruction, since there are no first order deformations of
$f\in\mgdd(W_r)$ along the fiber direction of $W_r\to D_r$, the
obstruction classes to extending families in $\mgdd(W_r)$ are exactly
the obstruction classes to extending them as families in $\mgdd(D_r)$.
Furthermore, the obstruction spaces (sheaves) to extending families in
$\mgdd(W_r)$ will then be the direct sum of the obstruction spaces
(sheaves) to extending them in $\mgdd(D_r)$ with the obstruction
spaces (sheaves) to deforming them along the fiber directions of
$W_r$. Let $V$ be the obstruction bundle to extending families along
the fiber directions of $W_r$.  According to the theory of virtual
moduli cycles (Li--Tian \cite{LT1,LT2}), the virtual moduli cycle of
$\mgdd(W_r)$ is the top Chern class of the obstruction bundle $V$
paired with the virtual moduli cycle of $\mgdd(D_r)$.

We now derive this obstruction bundle $V$.
Let $(f_r,\Sigma_r)\in\mgdd(W_r)$ be any relative map.
We let $(f,C)$ be the canonical extension of $(f_r,\Sigma_r)$ to $C\to\Po$
constructed before.
For any $r\pri\geq r$ we let $\Sigma_{r\pri}=C\cap f\upmo(D_{r\pri})$.
Then $f_{r\pri}=f|_{\Sigma_{r\pri}}\mh\Sigma_{r\pri}\to D_{r\pri}$ is
an element in $\mgdd(D_{r\pri})=\mgdd(W_{r\pri})$. In the following we
will identify the obstruction space $V(f_{r})$ to deforming  $f_{r}$
along the fiber directions of $W_{r}$. As expected, the space $V(f_r)$
will be canonically isomorphic to 
$V(f_{r\pri})$. 

We now derive the space $V(f_r)$. We first consider the case where
$f=1$ and $C$ is smooth. We pick a sufficiently small $\epsilon>0$ so
that $U_{r,1}=f\upmo(D_r^{\epsilon})\sub \Sigma_r$ 
is isomorphic to $\{r\upmo\leq |z|< r\upmo+\epsilon\}$ and that
$f|_{U_{r,1}}\mh U_{r,1}\to \bC$ is defined by $w=z^{-d}$.
Now we let $U_{r,0}=\Sigma_r-\partial\Sigma_r$. Then $U_{r,0}$ and
$U_{r,1}$ are an open covering of $\Sigma_r$ 
with
$$U_{r,1}\cap U_{r,0}\cong\{r\upmo<|z|<r\upmo+\epsilon\}.
$$
Our first task is to describe the deformations of $f_r|_{U_{r,0}}$ and
$f_r|_{U_{r,1}}$ along the fiber directions of $W_r$. Clearly,
deformations of $f_r|_{U_{r,0}}$ are given by the space of sections
$$s_0\in\Gamma(U_{r,0}, f_r\sta L_1\oplus f_r\sta L_2),
$$
whereas  deformations of $f_r|_{U_{r,1}}$ are given by the sections
$$s_1\in\Gamma(U_{r,1}, f_r\sta L_1\oplus f_r\sta L_2)
$$
subject to the boundary condition \ref{10.1}. We now investigate the
boundary condition in detail. Using the distinguished basis $e_1$ and
$e_2$, we can express $s_1$ as 
$$s_1=u_1(z) f_r\sta e_1+ u_2(z) f_r\sta e_2,
$$
where $u_1$ and $u_2$ are continuous functions over $U_{r,1}$ that are
holomorphic in the 
interior of $U_{r,1}$. We let $u_1(z)=\sum_{k=-\infty}^{\infty} a_k z^k$ and
$u_2(z)=\sum_{k=-\infty}^{\infty} b_k z^k$
be the Laurent series expansions of $u_1$ and $u_2$. Then the boundary
condition \ref{10.1} 
(after substituting $z$ by $r\upmo e^{i\theta}$)
is equivalent to
$$\sum_{k=-\infty}^{\infty} a_k r^{-k} e^{ik\theta}
=(r\upmo e^{-i\theta})^{-d}\sum_{k=-\infty}^{\infty} \bar b_k r^{-k} e^{-ik\theta},\
\theta\in\RR.
$$
This forces
\begin{equation}
\label{10.6}
a_k r^{-k}=\bar b_{d-k}r^k
\end{equation}
for all $k\in\ZZ$.
When $r\to\infty$, the above relations reduce to
\begin{equation}
\label{10.9}
  a_k=0\ \text{for}\ k<0;\quad b_k=0\ \text{for}\ k<d\and a_0=\bar b_d.
\end{equation}
We let
\begin{equation}
\label{10.21}
\Gamma(U_{r,1},f_r\sta L_1\oplus f_r\sta L_2))_r
=\{s\in \Gamma(U_{r,0},f_r\sta L_1\oplus  f_r\sta L_2)\mid s\ \text{satisfies}\ 
\ref{10.6}\ \},
\end{equation}
and
\begin{equation}
\Gamma(U_{r,1},f_r\sta L_1\oplus f_r\sta L_2)_{\infty}
=\{s\in \Gamma(U_{r,1},f_r\sta L_1\oplus f_r\sta L_2)\mid s\ 
\text{satisfies}\ \ref{10.9}\ \}
\end{equation}
According to the receipt of the obstruction theory, the obstruction space to
deforming
$$f_r\mh (\Sigma_r,\partial\Sigma_r)\lra (W_r)
$$
along the fiber directions of $W_r$ is the cokernel of
\begin{eqnarray}
\label{10.11}
&\Gamma(U_{r,1},f_r\sta L_1\oplus f_r\sta L_2)_{r}\oplus
\Gamma(U_{r,0},f_r\sta L_1\oplus f_r\sta L_2) 
\lra\\ 
&\hspace{3in}\Gamma(U_{r,0}\cap U_{r,1},f_r\sta L_1\oplus f_r\sta L_2).\nonumber
\end{eqnarray}
Using the relation \ref{10.6} it is easy to see that the above
cokernel is canonically isomorphic to 
the cokernel of
\begin{eqnarray}
\label{10.24}
&\Gamma(U_{r,1},f_r\sta L_1\oplus f_r\sta L_2)_{\infty}\oplus
\Gamma(U_{r,0},f_r\sta L_1\oplus f_r\sta L_2) 
\lra\\
&\hspace{3in} \Gamma(U_{r,0}\cap U_{r,1},f_r\sta L_1\oplus f_r\sta L_2).\nonumber
\end{eqnarray}
Our next step is to repress this cokernel in terms of some cohomology
groups of line bundles over $C$. 
For this we need to extend the line bundles $f_r\sta L_1$ and $f_r\sta L_2$
to line bundles over $C$. We extend both line bundle to trivial line
bundles over $C$ 
so that the constant sections $e_i$ of $L_i$ pull back to a constant
section of $\cO_C$. 
With this choice of extensions, $\Gamma(U_{r,1},f_r\sta L_1\oplus
f_r\sta L_2)_{\infty}$ then is canonically isomorphic to
\begin{equation}
\label{10.22}
\Gamma(\cO_{\cU_{1}}(-x))\oplus\Gamma(\cO_{\cU_{1}}(-dx)).
\end{equation}
Here $\cU_1$ is the disk $\{|z|<\epsilon\}\sub C$, $x\in \bar U_{r,1}$ is the
point ramified over $\qi\in\Po$ and $\cU_0=C-x$.
Hence the cokernel of \ref{10.11} becomes the cokernel of
\begin{eqnarray}
\label{10.12}
&\Gamma(\cU_1,\cO_{C}(-x)\oplus \cO_{C}(-dx))
\oplus \Gamma(\cU_0,\cO_{C}(-x)\oplus \cO_{C}(-dx))
\lra\\
&\hspace{3in}  \Gamma(\cU_0\cap \cU_1,\cO_{C}(-x)\oplus \cO_{C}(-dx)),\nonumber
\end{eqnarray}
which by the definition of \v Cech cohomology is
\begin{equation}
\label{10.13}
H^1(\cO_C(-x))\oplus H^1(\cO_C(-dx)).
\end{equation}
This shows that the obstruction space $V(f_r)$ of $f_r\in\mgdd(W_r)$
is the cohomology group \ref{10.13}.

The vector spaces $V(f_r)$ over $f_r\in\mgdd(W_r)$ form a vector
bundles over $\mgdd(W_r)$. 
Since $V(f_r)$ is independent of $r$, this vector bundle extends to
$\mgdd(\Po)_0$ via the same formula. We denote this vector bundle over
$\mgdd(\Po)_0$ by $V$. 

\begin{prop}
\label{ob}
Let $f_r\in\mgdd(\Po)$ be any relative stable morphism with $C$ the domain of
$f$ and let $D_\bd=d_1x_1+\cdots+d_hx_h$ be the distinguished divisor in $C$.
We let $D=x_1+\cdots+x_h$ and let  $V$ be the vector bundle over
$\mgdd(\Po)$ whose fibers over $f$ are 
$$V(f)=H^1(\cO_C(-D)\oplus\cO_C(-D_\bd)). 
$$
Then the obstruction bundle to deforming maps in
$\mgdd(W_r)$ along the fiber directions of $W_r$ is given by the vector bundle
$V|_{\mgdd(W_r)}$.
\end{prop}

\begin{proof}
The proof is exactly the same as the case $h=1$ and $\Sigma_r$ is
smooth, except that, in deriving this cohomology space, we need to cover
$\Sigma_r$ by more than two open subsets. 
\end{proof}

According to the general principle of virtual moduli cycles, the
expected (virtual) number of maps in $\mgdd(W_r)$ should be 
\begin{equation}
\label{int}
\int_{[\mgdd(W_r)]\vir} c_{\text{top}}(V).
\end{equation}
Here $[\mgdd(W_r)]\vir$ is the virtual moduli cycle of $\mgdd(W_r)$
and $c_{\text{top}}$ is the 
top Chern class of a vector bundle.
Clearly, this integration is not well defined mathematically
because $\mgdd(W_r)$ is not compact. To make sense of this integral we
need to compactify $\mgdd(W_r)$ and then extend $V$.
Since $\mgdd(W_r)$ is naturally an open subset of $\mgdd(\Po)$, we
can use $\mgdd(\Po)$ to compactify $\mgdd(\Po)_0$. It was proved by
the first author of the present paper that 
$\mgdd(\Po)$ is a proper Deligne--Mumford stack, and that it admits a perfect
obstruction theory and thus comes with a natural choice of virtual moduli
cycles, denoted by $[\mgdd(\Po)]\vir$. 

Extending $V$ to $\mgdd(\Po)$ needs more work.
There is an obvious extension as follows: Let $f\in\mgdd(\Po)$ be any relative
stable morphism with domain $C$ and distinguished divisor $D_\bd$ as
before. The the 
vector bundle $\tilde V$ whose fiber over $f$ is the vector space
$V(f)$, is a vector bundle over $\mgdd(\Po)$. However, there probably
are other extensions, and at the 
moment we have no reason as to which one is a natural choice of
extension. Note that different extensions may give different numerical
answers to the integral. 
The choice of the right extension requires a detailed analysis of the
obstruction theory of $f$ near the boundary of $\mgdd(W_r)$, and we
will address this issue in our future research.


\noindent
\section{Relative stable morphisms and localization}
\label{s:localization}

In this and the next sections, we will use  localization to evaluate
the integral 
\ref{int}.   In \fullref{s:modulispace}, we defined an $S^1$--action on $W_r$ via the rule
$$(w,u,v)^t=(t\upmo w, tu, v),\quad t=e^{2\pi i\theta}\quad \text{where}\ \theta\in\RR.
$$
This action extends to the total space $\bone_{\bC}\oplus\bone_{\bC}$
via the same rule. As we mentioned before, this action preserves the boundary
condition $B_r\sub W_r$, and thus
it is reasonable to expect that localization using this $S^1$--action
will give us the correct answer.

Let $\mgdd(\Po)_0$ be the moduli space of ordinary relative stable
morphisms defined 
before. The given $S^1$--action on $W_r$ induces a natural $S^1$--action
on $\mgdd(\Po)_0$. 
In this section, we will study the fixed loci of the
induced $S^1$--action on $\mgdd(\Po)_0$ and its related information
required in the later computation. Let $f\in\mgdd(\Po)_0$ be any
relative stable map. 
As before, we denote by $(C,D_\bd)$  the pair of the domain of $f$ and
the distinguished divisor of $f$.
Since $f\in\mgdd(\Po)_0$, the codomain of $f$ is $\Po$ and
$f\upmo(\qi)=D_\bd$. 
We recall that if we denote by $g_t$ the $S^1$--action then
$$g_t  [w,1] = [tw,1]\quad\text{and}\quad g_{t*}(w) = g^*_{t^{-1}}w=t^{-1}w.
$$
If we use $\wt$ to denote the weight of the $S^1$--action, then the
function $w$ has weight $-\wt$. 
 The two fixed points of the
$S^1$--action on $\IP^1$ are $q_{\mtiny{0}} =[0,1] \in \IP^1$ and
$q_\infty=[1,0]\in\IP^1$.

In the above notation, considering source Riemann surfaces with
one hole corresponds to setting $h=1$, and in the remainder of
this section that is what we will do.  For genus $g=0$, there is
only one fixed point in $\mgdd(\Po)_0^{S^1}$ given by the map
$$f\co  \Po\to \Po, \quad f([z,1])=[z^d,1].
$$
For $g>0$, the fixed loci are the image of the embedding
$$\bM_{g,1}\lra \mgdd(\Po)_0^{S^1}
$$
that sends any $(C_2,p)\in\bM_{g,1}$ to the relative stable morphism
$$f\co  (C, dx)\lra \Po
$$
defined as follows: First, the curve $C$ is the gluing of $\Po$ with
$C_2$ along the points $0\in\Po$ and $p\in C_2$; the restriction of
$f$ to $C_2$ is the constant map sending $C_2$ to $\qz$ and the
restriction of $f$ to $C_1\equiv\Po$ is define by $[z,1]\mapsto
[z^d,1]$. 
Since $w=z^d$ and the weight of $w$ is $-\wt$, the weight of the
function $z$ is  $-\wt/d$. For simplicity, in the following we will
simply denote $f|_{C_i}$ by 
$f_i$. We will use $p$ to denote the node in $C$.

If we consider the full moduli space $\mgdd(\Po)$, however, there are
other fixed loci of the ${S^1}$--action.  As described in the previous
section, a relative stable morphism $f\in\mgdd(\Po)$ is a morphism
$f\co C\rightarrow\IP^1[m]$. The ${S^1}$--action extends to $\IP^1[m]$
an ${S^1}$--action on $\mgdd(\Po)$. We will denote the fixed loci of
$\mgdd(\Po)$ contained in $\mgdd(\Po)_0$ by $\Theta_I$. The rest of
the ${S^1}$--fixed loci of 
$\mgdd(\Po)$ will be denoted by $\Theta_{II}$.

Now let $N_{\Theta_I}$ be the equivariant normal bundle to the fixed
loci $\Theta_I$.  In this part, we will compute the
equivariant Euler class $e(N_{\Theta_I})$ of $\Theta_I$.
Let $(f,C,D_\bd)\in \Theta_I$ be any point. The
tangent space and the obstruction space to deforming $f$ are given by the
extension groups
$$
T^\bullet\defeq \Ext_C^\bullet([f^*\Om_{\IP^1} (\log q_\infty) \rightarrow \Om_C(x)], \ocal_C) ,
$$
which fit into the long exact sequence
\begin{eqnarray}
&0& \lra \Ext_C^0(\Om_C(x),\ocal_C) \lra
\Ext_C^0(f^*\Om_{\IP^1}(\log
q_\infty),\ocal_C) \lra T^1 \lra \nonumber\\
&&\lra \Ext_C^1(\Om_C(x),\ocal_C) \lra
\Ext_C^1(f^*\Om_{\IP^1}(\log q_\infty), \ocal_C) \lra T^2 \lra 0 .
\label{eq:T-O}
\end{eqnarray}
From this information, we can obtain the equivariant Euler class
$e(N_{\Theta_I})$. Following the notation of Graber and
Pandharipande \cite{GP}, it is given by
\begin{equation}
e(N_{\Theta_I}) = {e(B_{II}^m) e(B_{IV}^m) \over e(B_I^m)
e(B_{V}^m)}, \label{eq:EulerN}
\end{equation}
where $B^m_i$ denotes the moving part of the $i^{\rm th}$ term in the
sequence \eqr{eq:T-O}.

We now consider the case $g>0$. In this case $C=C_1\cup C_2$ with the node $p$.
We let $i_1\mh C_1\ra C$ and $i_2\mh C_2 \ra C$ be inclusion maps. Then
$$
\Om_C =  {i_1}_* \Om_{C_1} \oplus {i_2}_* \Om_{C_2} \oplus \IC_p.
$$
Then
\begin{eqnarray}
\Ext_C^0(\Om_C(x),\ocal_C) &=& \Hom_C (\iof\Om_{C_1} (x), \ocal_C)
\oplus \Hom_C(\itf \Om_{C_2} (x), \ocal_C)\nonumber\\
&&\hspace{2.5in} \oplus
\Hom_C(\IC_p,\ocal_C) \nonumber
\\
&=& \Hom_{C_1} (\Om_{C_1}(x), \ocal_{C_1} (-p)) =
\mbox{H}^0_{C_1}(T_{C_1}(-p-x)). \nonumber
\end{eqnarray}
It is easy to see that the second arrow in \ref{eq:T-O} is injective.
Thus we can take $e(B_I^m)=1$ while keeping track of the term in $e(B_{II}^m)$
to be cancelled.

As to the term $e(B_{II}^m)$, we notice that
$$
\Ext_C^0(f^*\Om_{\IP^1}(\log q_\infty),\ocal_C) =
\mbox{H}^0_{C_1}(f^*_1 T_{\IP^1}(-d\cdot x)).
$$
$\mbox{H}^0_{C_1}(f^*_1T_{\IP^1}(-dx))$ has dimension $d+1$ and
its basis is given by
$$
\left\{ \pdd{w}, z \pdd{w}, z^2 \pdd{w}, \ldots,
z^{d-1}\pdd{w},z^{d}\pdd{w} \right\}
$$
with weights
$$
\left\{ \wt\,,  \frac{d-1}{d}\,\wt\,,  \frac{d-2}{d}\,\wt\,,\ldots,
\frac{1}{d}\,\wt\,,0 
\right\}.
$$
Thus, modulo the weight-zero piece which was cancelled by the second
arrow in \ref{eq:T-O}, we have
\begin{equation}
\frac{e(B_{II}^m)}{e(B_I^m)}=
\prod_{j=0}^{d-1} {d-j \over d} \, \wt = {d!\over d^d}\,
\wt^d. \label{eq:eII} 
\end{equation}
We now look at the term $e(B_{IV}^m)$. We have
\begin{eqnarray}
\Ext_C^1(\Om_C(x),\ocal_C) &=& \Ext_C^0(\ocal_C, \Om_C(x)\otimes
\om_{\mbox{\tiny $C$}})^\vee \nonumber
\\
&=&\Ext^0_{C_2} (\ocal_{C_2}, \om_{\mbox{\tiny $C_2$}}^{\otimes
2}(p))^\vee \oplus \Ext^0_{C}(\ocal_C,\IC_p\otimes\om_{\mbox{\tiny
$C$}})^\vee \nonumber
\\
&=&\Ext^1_{C_2}(\Om_{\mbox{\tiny $C_2$}}(p),\ocal_{C_2})\oplus
 T_{C_1,p}\dual\otimes T_{C_2,p}^\vee.  \nonumber
\end{eqnarray}
Since $f(C_2)=q_0$,
$\Ext^1_{C_2}(\Om_{C_2}(p),\ocal_{C_2})$ lies in the
fixed part of $\Ext_C^1(\Om_C(x),\ocal_C)^{S^1}$.
On the other hand, $T_{C_1,p}\dual$ has weight $\wt/d$ while
$T_{C_2,p}\dual$ has weight $0$. Hence
\begin{equation}
e(B_{IV}^m) = {1 \over d}\, \wt- \psi = {\wt- d\cdot\psi \over
d},\label{eq:eIV} 
\end{equation}
where $\psi=c_1({\cal L}_p)$ and ${\cal L}_p \to
\overline{M}_{g,1}$ is the cotangent line bundle whose fiber over $(C_2,p)$ is
$T_{C_2,p}\dual$.

Finally, it is direct to compute that
\begin{eqnarray}
 \Ext^1_C(f^*\Om_{\IP^1}(\log
q_\infty), \ocal_C)
= \mbox{H}^1_{C_2} (\cO_{C_2})\otimes T_{q_0} \IP^1.
\end{eqnarray}
Note that $\mbox{H}_{C_2}^1(\ocal_{C_2})$ gives the dual of the Hodge bundle
$E$ on $\overline{M}_{g,1}$.  Hence $e(B_V^m)$ is the top Chern class
of the bundle $ E\dual\otimes T_{q_0}\IP^1$ .
Since the representation of the $S^1$--action induced on $T_{q_0}\IP^1$ has
weight $\wt$, we obtain
\begin{eqnarray}
e(B_V^m)&=&c_{\text{top}}(E\dual\otimes T_{q_0} \IP^1)\\
&=& \left(\wt^{g}+c_1(E^\vee)\,\wt^{g-1}+c_2(E^\vee)\,\wt^{g-2} + \cdots +
c_g(E^\vee)\right).
\end{eqnarray}
Hence, we arrive at the following result:
\begin{eqnarray}\label{eq:EulerNg}
&\hspace{-2in}{1\over e(N_{\Theta_I})} = {d^d\over d!} \, \wt^{-d}\left( {d\over \wt
- d\cdot\psi}\right)\\
&\hspace{1in} \left(\wt^{g} + c_1(E^\vee)\, \wt^{g-1} +
c_2(E^\vee)\, \wt^{g-2} + \cdots + c_g(E^\vee)\right)\,. \nonumber 
\end{eqnarray}
In the case $g=0$, a similar computation yields
\begin{equation}
{1\over e(N_{\Theta_I})} =  {1\over d}\cdot {d^d\over d!}\, \wt^{1-d}\, .\label{eq:EulerN0}
\end{equation}


\section{Localization of the integral}
\label{s:RelGWh=1}

We now evaluate the contribution of the integral
\begin{equation}
\int_{[\mgdd(\Po)]\vir} c_{\mbox{\scriptsize  top}} (V)
\label{eq:integral}
\end{equation}
at the fixed loci $\Theta_I$ using the $S^1$--action
given before. Here $V$ is the
obstruction vector bundle defined in \fullref{ob}.
Using the localization theorem
\cite{GP}, we have
$$
\left[\int_{[\mgdd(\Po)]\vir} c_{\mbox{\scriptsize  top}} (V) \right]_{\Theta_I}
= \left[{1\over |\mbox{A}_{\Theta_I}|} \int_{\Theta_I} {\iota\sta
(c_{\mbox{\scriptsize  top}} (V)) \over e(N_{\Theta_I})}\right]_0,
$$
where $\iota$ is the inclusion
${\Theta_I}\to \mgdd(\Po)$ and $\mbox{A}_{\Theta_I}$ is a
automorphism group defined as in \cite{GP}.

We notice that by Riemann--Roch theorem,
\begin{eqnarray}
\mbox{dim}_{\mscript{\IC}}\, \mgdd(\Po) & =& 2d +
(1-g)(\mbox{dim}_{\mscript{\IC}}\, \IP^1 - 3) - (\mbox{deg}\,
(\bd) -
\ell (\bd)) \nonumber \\
&=& 2g-2+h+d\nonumber \\
&=& \mbox{dim}_{\mscript{\IC}}
\mbox{$\mbox{H}^1(C,\calo_C(-D_\bd)\oplus {\cal
O}_C(-D))$}. \nonumber
\end{eqnarray}
In this section, we focus on source Riemann surfaces with one hole
$(h=1)$,  in which case we need to find the equivariant Chern class
$c_{\text{top}}(V)$ where $V$ is defined as follows.  Let $f\in \Theta_I$ be
a point as before. Then the fiber of $V$ over $f$ is
$$H^1(\cO_C(-x))\oplus H^1(\cO_C(-dx)).
$$
For simplicity, we denote by $V_1$ the subbundle of $V$ whose fibers over $f$
are the vector space $H^1(\cO_C(-x))$, and by $V_2$ the subbundle of $V$ whose
fibers over $f$ are $H^1(\cO_C(-dx))$.
The line bundle $\cO_C$ in the first
cohomology group has 
weight $-\wt$ while the line bundle $\cO_C$ in the second cohomology group has
weight $0$. We will work out the
genus-zero and higher genus cases
separately.

We first look at the genus $g=0$ case.
In this case, $\mbox{H}^1(C,\calo_{C}(-x))=0$ and hence $V=V_2$.
To analyze the contribution from $\mbox{H}^1(C,\calo_{C}(-dx))$, we use
the exact sequence 
$$
0 \lra \calo_C (-dx) \lra \calo_C \lra \calo_{dx} \lra 0
$$
and the induced  cohomology exact sequence
$$
0 \lra \mH^0(\calo_C) \lra \mH^0 (\calo_{dx}) \lra \mH^1(\calo_C
(-dx)) \lra 0.
$$
A basis of $\mH^0(\calo_C)$ is just $\{1\}$ and that of $\mH^0
(\calo_{dx})$ is $\{1,z^{-1},z^{-2},\ldots,z^{-(d-1)}\}$.
Hence a basis of
$\mH^1(\calo_C(-dx))$ is
$$
\left\{\frac{1}{z},\frac{1}{z^2},\ldots,\frac{1}{z^{d-1}}\right\}
$$
with weights
\begin{equation}
\left\{ \frac{1}{d}\wt\,,\frac{2}{d}\wt\,,\ldots,\frac{d-1}{d}\wt \right\}.
\label{eq:WH1C1O-dx}
\end{equation}
Thus the top Chern class of the obstruction
bundle $V$ is
\begin{equation}
c_{\text{top}}(V)=c_{\text{top}}(V_2)=
\prod_{j=1}^{d-1} \frac{j}{d}\, \wt = \frac{(d-1)!}{d^{d-1}}\,
\wt^{d-1}. \label{eq:OB0} 
\end{equation}
In the cases with $g>0 $, $C$ is a union of the
two irreducible components $C_1=\IP^1$ and $C_2=\Sigma_g$
intersecting at the node $p$.  Then there is the normalization exact
sequence
\begin{equation}
0 \lra \calo_C (-d x) \lra \calo_{C_1}(-dx) \oplus \calo_{C_2}
\lra \calo_p \lra 0,
\end{equation}
which gives the following exact sequence of cohomology:
\begin{eqnarray*}
&0\lra \mH^0_{C_2}(\calo_{C_2}) \stackrel{\cong}{\lra}
\mH^0_C(\calo_p)\lra \mH^1_C(\calo_C(-dx))
\lra\\
&\hspace{2in}\mH^1_{C_1}(\ocal_{C_1}(-dx))\oplus \mH^1_{C_2}(\calo_{C_2}) \lra
0.
\end{eqnarray*}
So $\mH^1(C,\calo_C(-dx))$ is given by
\begin{equation}
\mH^1(C,\calo_C(-dx)) =  \mH^1(C_1,\ocal_{C_1}(-dx))\oplus
\mH^1(C_2,\calo_{C_2}).
\label{eq:H1O-dx}
\end{equation}
We have already computed the contribution of
$\mH^1(C_1,\calo_{C_1}(-dx))$ to the equivariant Chern class
$c_{\text{top}}(V_2)$.  Since
$\calo_{C_2}$ in \eqr{eq:H1O-dx} has weight zero and there is no
twisting, the contribution from
$\mH^1(C_2,\calo_{C_2})$ to $c_{\text{top}}(V)$ is
$c_g(E^\vee)=(-1)^g c_g(E)$.
Hence the contribution from
$\mH^1(C,\calo_C(-dx))$ to the equivariant top Chern class $c_{\text{top}}(V_2)$ is
$$
(-1)^g {(d-1)! \over d^{d-1}} c_g(E)\, \wt^{d-1} .
$$
In a similar spirit, we can use the same line of reasoning to get
\begin{equation}
\mH^1(C,\calo_C(-x)) =  \mH^1(C_1,\ocal_{C_1}(-x))\oplus
\mH^1(C_2,\calo_{C_2}).
\label{eq:H1O-x}
\end{equation}
As discussed in the genus-zero case, $\mH^1(C_1,\ocal_{C_1}(-x))$
has dimension zero and does not contribute.  To compute the
contribution from $\mH^1(C_2,\calo_{C_2})$, we recall that, as
mentioned in the beginning of this section, $\calo_{C_2}$ in the
present case has weight $-\wt$.  Hence
\begin{eqnarray}
c_{\text{top}}(V_1)&=& (-1)^g
\left(\wt^g-c_1(E^\vee)\, \wt^{g-1}+c_2(E^\vee)\, \wt^{g-2}+\cdots+
(-1)^g c_g(E^\vee)\right)\nonumber \\
&=& (-1)^g\left(\wt^g + c_1(E) \wt^{g-1} +c_2(E)\wt^{g-2} + \cdots+ c_g(E)\right).
\end{eqnarray}
In summary, we have
\begin{equation}
c_{\text{top}}(V)=
{(d-1)! \over d^{d-1}}\, c_g(E)
\left(\wt^g+c_1(E)\, \wt^{g-1}+c_2(E)\,
\wt^{g-2}+\cdots+c_g(E)\right)\, \wt^{d-1} .
\label{eq:WH1OBg}
\end{equation}
We are now ready to evaluate the contribution from $\Theta_I$ to the
integral \ref{eq:integral}
in the case $h=1$. We begin with the case $g=0$.

After taking into account the automorphism
group $A_{\Theta_I}$, which has order $d$, the genus-zero answer
is given by multiplying the expressions in \eqr{eq:EulerN0} and
\eqr{eq:OB0}.  This gives the virtual number in the genus-zero case
invariant
\begin{equation}
{1 \over d}\cdot
{1\over d}{d^d\over d!}\, \wt^{1-d}\,\cdot {(d-1)!\over d^{d-1}}\,
\wt^{d-1} = {1\over d^2}, 
\label{eq:GW01}
\end{equation}
which agrees with \eqr{eq:Fgh}.

We now consider the cases $g>0$.
We need to use the results in
\eqr{eq:EulerNg} and \eqr{eq:WH1OBg} to get the higher genus
invariants. It follows from Mumford's formula
$$
(1+c_1(E^\vee)+c_2(E^\vee)+\cdots+c_g(E^\vee)) \cdot
(1+c_1(E)+c_2(E)+\cdots+c_g(E)) = 1
$$
that
\begin{eqnarray*}
&\hspace{-2in}(\wt^g+c_1(E^\vee)\, \wt^{g-1}+c_2(E^\vee)\,
\wt^{g-2}+\cdots+c_g(E^\vee)) \cdot\\ 
&\hspace{1.5in}
(\wt^g+c_1(E)\, \wt^{g-1}+c_2(E)\, \wt^{g-2}+\cdots+c_g(E)) = \wt^{2g}.
\end{eqnarray*}
Hence we have
\begin{eqnarray}
\left[\int_{[\mgdd(\Po)_0]\vir} c_{\text{top}}(V)\right]_{\Theta_I}
&=& \left[ {1\over |\mbox{A}_{\Theta_I}|} \int_{\overline{M}_{g,1}}
{d\,\cdot \wt^{2g-1} \over 
\wt-d\cdot\psi} c_g(E)\right]_0\nonumber \\
&=& d^{2g-2}  \int_{\overline{M}_{g,1}} \psi^{2g-2} c_g(E),
\end{eqnarray}
where in the last equality we have used the fact that the moduli
space $\overline{M}_{g,1}$ of Deligne--Mumford stable curves has
dimension $\mbox{dim}_{\mscript{\IC}}\overline{M}_{g,1}=3g-2$. The
Hodge integral above can be easily evaluated by using Faber  and
Pandharipande's generating function for Hodge integrals over the
moduli space ${\overline{M}_{g,1}}$ \cite{FP}. Taking the result
from \cite{FP}, we conclude that
\begin{equation}
\left[\int_{[\mgdd(\Po)_0]\vir} c_{\text{top}}(V)\right]_{\Theta_I}= d^{\,2g-2}
{2^{2g-1}-1\over 2^{2g-1}} {|B_{2g}| \over (2g)!}.\label{eq:GWg1}
\end{equation}
As promised, \eqr{eq:GWg1} is precisely equal to the expected answer
\eqr{eq:Fgh}.


\section{Invariants for $h>1$}
\label{s:RelGWh>1}

As mentioned before, all invariants for $h>1$
vanish.  The main idea that underlies our argument is that at least
one of the weights of the $S^1$--action on the 
obstruction bundle is zero. 
We will present our argument for genus-zero and higher genus cases
separately.  

We will first consider the genus-zero case.
Assume that $h=2$, in which case $\mu=(d_1,d_2)$, where $d_1+d_2=d$.
In genus-zero $C=C_1 \sqcup_p C_2$, where $C_1$ and
$C_2$ both are rational curves and $p$ is a node that gets mapped to $\qz$.
For $i=1 {\rm\ or\ } 2$, $d_i>0$ is the degree of the map $f_i$ that
maps $C_i$ to $\IP^1$. 
If we denote the pre-images of $q_\infty$ by
$x_1\in C_1$ and $x_2\in C_2$, then we have the normalization exact
sequence 
\begin{eqnarray*}
&\hspace{-1in}0 \lra \calo_C (-d_1 x_1 - d_2 x_2) \lra \calo_{C_1}(-d_1x_1) \oplus
\calo_{C_2}(-d_2x_2) \lra \\
&\hspace{2in}
\calo_C (-d_1x_1-d_2x_2)\vrule\,_p \lra 0,
\end{eqnarray*}
which gives the long exact sequence of cohomology
\begin{eqnarray}
&\hspace{-.5in}0\lra \mH^0(C,\calo_C(-d_1x_1-d_2x_2)\vrule\,_p) \lra
\mH^1(C,\calo_C(-d_1x_1-d_2x_2)) \lra \nonumber \\
&\hspace{1.3in}\lra \mH^1(C_1,\ocal_{C_1}(-d_1x_1))\oplus
\mH^1(C_2,\calo_{C_2}(-d_2x_2)) \lra 0. \nonumber
\end{eqnarray}
From this we immediately see that one of the $S^1$--action weights on
$\mH^1(C,\calo_C(-d_1x_1-d_2x_2))$ is zero, since the
weight on  $\mH^0(C,\calo_C(-d_1x_1-d_2x_2)\vrule\,_p)$ 
is zero.  This means that the contribution 
of $\mH^1(C,\calo_C(-d_1x_1-d_2x_2))$ to
the equivariant top Chern class of the
obstruction bundle vanishes,  thus rendering the invariant to vanish
as well. 

For $h=3$, $C$ contains a contracted genus-zero component
$\tilde{C}_0$ which is 
connected to $3$ rational curves, say $C_1,C_2,C_3$, at $3$ nodes,
say $p_1,p_2,p_3$.  Note that since $\tilde{C}_0$ contains 3 special
points, it is stable and can be contracted to $\qz$.  Each $C_i$ maps
to $\IP^1$  with degree $d_i>0$ and contains a special point $x_i$
that gets mapped to $q_\infty$.  
As usual there is the exact normalization sequence
$$
0 \lra \calo_C (-\sum_i^3d_i x_i) \lra \bigoplus_{i=1}^3
\calo_{C_i}(-d_ix_i) \oplus \calo_{\tilde{C}_0} \lra 
\bigoplus_{i=1}^3 \calo_C (-\sum_i^3d_i{ x_i)\vrule\,_{p_i}} \lra 0,
$$
and the associated cohomology long exact sequence
\begin{eqnarray}
&&\hspace*{0in} 0\lra \mH^0(\tilde{C}_0,\calo_{\tilde{C}_0})\nonumber \\
&&\hspace*{1in}  \stackrel{\psi}  \lra
      \bigoplus_i^3 \mH^0(C,\calo_C(-\sum_i^3 {d_i x_i)\vrule\,_{p_i}})
 \lra 
 \mH^1(C,\calo_C(-\sum_i^3d_i x_i) \lra \nonumber \\
&&\hspace*{2in}\lra \bigoplus_{i=1}^3 \mH^1(C_i,\ocal_{C_i}(-d_ix_i))\oplus
\mH^1(\tilde{C}_0,\calo_{\tilde{C}_0}) \lra 0.\nonumber
\end{eqnarray}
Here, $\psi$ is not surjective and it follows that not all of the
zero weights on 
$$\bigoplus_i^3 \mH^0(C,\calo_C(-\sum_i^3 {d_ix_i)\vrule\,_{p_i}})$$ get
cancelled in
\begin{eqnarray}
\mH^1(C,\calo_C(-\sum_i^3d_i x_i)) &=&
\bigoplus_{i=1}^3 \mH^1(C_i,\ocal_{C_i}(-d_ix_i))\oplus
\mH^1(\tilde{C}_0,\calo_{\tilde{C}_0})
- \mH^0(\tilde{C}_0,\calo_{\tilde{C}_0})\nonumber\\
&&\hspace{0.5in} +\, \bigoplus_i^3 \mH^0(C,\calo_C(-\sum_i^3 {d_i
x_i)\vrule\,_{p_i}}) . \nonumber 
\end{eqnarray}
Therefore, the equivariant top Chern class of
the obstruction bundle in the localization theorem vanishes.  We can 
perform induction on $h$ and 
conclude that genus-zero invariants vanish
for all $h>1$.  We will now sketch how that works. 
For $h\leq n$, assume that the $S^1$--action on
$\mH^1(C,\calo_C(-\sum_i^h d_i x_i))$ has at least one zero weight
and that therefore the invariants vanish.  At
$h=n+1$, a $S^1$--fixed stable map  can be constructed from that at
$h=n$ by attaching a 
rational curve $C_{n+1}$ to the contracted component, 
such that  
$\mbox{deg}(f|_{C_{n+1}}) = d_{n+1} > 0$. $C_{n+1}$ contains the point
$x_{n+1}$ that gets mapped to $q_\infty$ and is joined to the
contracted component at a new node.  
Such an operation increases  the number of nodes  by 1,
and analyzing the exact normalization sequence and its
associated cohomology long exact sequence shows that
the number of zero weights on 
$\mH^1(C,\calo_C(-\sum_i^{h} d_i x_i))$ has increased by one. 
Therefore, the total  number of zero weights on
$\mH^1(C,\calo_C(-\sum_i^h d_i x_i))$ is again non-zero.  
This shows that the equivariant top Chern class of the obstruction
bundle vanishes at $h=n+1$.

Now assume that $g\geq 1$  and $h=2$.
In addition to the two rational curves $C_1$ and $C_2$, we 
introduce a stable genus--$g$ curve $\tilde{C}_g$, which gets contracted to 
$\qz$.  There are two nodes $p_1$ and $p_2$ where $C_1$ and $C_2$,
respectively, intersect $\tilde{C}_g$.  In our usual notation, the
normalization exact sequence in the present case is
\begin{eqnarray*}
&\hspace{-1in}0 \lra \calo_C (-d_1 x_1 - d_2 x_2) \lra \bigoplus_{i=1}^2
\calo_{C_i}(-d_ix_i) \oplus \calo_{\tilde{C}_g} \lra\\ 
&\hspace{2in}\bigoplus_{i=1}^2 {\calo_C (-d_1x_1-d_2x_2)\vrule\,_{p_i}} \lra 0.
\end{eqnarray*}
This implies the following long exact sequence of cohomology:
\begin{eqnarray}
&& 0\lra \mH^0(\tilde{C}_g,\calo_{\tilde{C}_g})\nonumber\\ 
&&\hspace{.2in}\stackrel{\varphi}{\lra}
      \mH^0(C,\calo_C(-d_1x_1-d_2x_2)\vrule\,_{p_1})\oplus 
       \mH^0(C,\calo_C(-d_2x_2-d_2x_2)\vrule\,_{p_2})
 \lra \nonumber \\
&&\hspace{.2in}\mH^1(C,\calo_C(-d_1x_1-d_2x_2))
\lra \bigoplus_{i=1}^2 \mH^1(C_i,\ocal_{C_i}(-d_ix_i))\oplus
\mH^1(\tilde{C}_g,\calo_{\tilde{C}_g}) \lra 0.\nonumber
\end{eqnarray}
Unlike in the $h=1$ case, $\varphi$ is not surjective and we need to compute
\begin{eqnarray*}
&
\hspace{-3in}\mH^1(C,\calo_C(-d_1x_1-d_2x_2))\\ 
&\hspace{.5in}=
\bigoplus_{i=1}^2 \mH^1(C_i,\ocal_{C_i}(-d_ix_i))\oplus
\mH^1(\tilde{C}_g,\calo_{\tilde{C}_g})
- \mH^0(\tilde{C}_g,\calo_{\tilde{C}_g}) + \\
&\hspace{.5in} +\, \mH^0(C,\calo_C(-d_1x_1-d_2x_2)\vrule\,_{p_1})\oplus 
       \mH^0(C,\calo_C(-d_2x_2-d_2x_2)\vrule\,_{p_2}). 
\end{eqnarray*}
The zero weight term from $\mH^0(\tilde{C}_g,\calo_{\tilde{C}_g})$ will cancel
only one of the two zero weight 
terms from the second line,
leaving a zero weight term in
\mbox{$\mH^1(C,\calo_C(-d_1x_1-d_2x_2))$}.   Hence the 
equivariant top Chern class of the obstruction bundle 
again vanishes, and so 
does the invariant.  

The vanishing of the invariants for $g\geq 1$ again follows from
induction on 
$h$.  As in the genus-zero case, a $S^1$--fixed stable map at $h=n+1$
can be constructed from that at $h=n$ by attaching a non-contracted
rational curve, say $C_{n+1}$, to the contracted component
$\tilde{C}_g$ at a new node.  
This addition of a node increases the number of zero weights of the
$S^1$--action on $\mH^1(C,\calo_C(-\sum_i^h d_ix_i))$, and therefore
the equivariant top Chern class of the obstruction
bundle vanishes at $h=n+1$ as it does at $h=n$.
Hence,  all higher genus invariants vanish for $h>1$.

To recapitulate, we have just established that 
$$
\left[\int_{[\mgdd(\Po)]\vir} c_{\mbox{\scriptsize  top}} (V)
\right]_{\Theta_I} 
= 
 0 \ , \ \ \ \forall g
\geq 0,\  d\geq h>1,
$$
in perfect agreement with what was expected from \sref{s:description}.

\section{Conclusion}
\label{s:Conc}
In this paper we have made an explicit connection between topological
open string theory and relative stable morphisms.  In the particular
example we consider, we have successfully reproduced open string
instanton multiple cover answers as invariants of relative stable
maps.  So far several interesting proposals for studying open string
instanton effects have been made (Ooguri--Vafa \cite{OV},
Kachru--Katz--Lawrence--McGreevy \cite{KKLM} and Aganagic--Vafa
\cite{AV}), but direct computational methods involving integrals over
moduli spaces of stable morphisms have been hitherto lacking.  This is
in marked contrast to the closed string case, where there exist
well-developed techniques in the context of Gromov--Witten theory
(Faber, Graber and Pandharipande \cite{GP,FP}).

Open string instantons  play an important role in string theory.  
For example, in the Strominger--Yau--Zaslow conjecture of mirror
symmetry, open string instanton effects are crucial for modifying the
geometry of D--brane moduli space \cite{SYZ}.  Also, genus-zero topological
open string amplitudes are important for computing superpotentials
in ${\cal N}=1$ supersymmetric theories in 4--dimension---see
\cite{KKLM,OV} and references therein.  
It is clear that many illuminating implications can stem from
understanding better how one can directly compute open string instanton
amplitudes. We hope we have made it clear in this paper
that relative stable morphisms 
could be the right framework for studying open string instantons in
general, and that the proposed link between topological open string
theory and relative stable maps well deserve further
investigations. 

Applying the theory of relative stable morphisms to topological open
string theory is in the incipient stage.  In a sense we have studied
in this paper what could be considered the simplest example.  As
mentioned in \sref{s:description}, the quantities we have reproduced
correspond to the invariants of a simple knot in $S^3$.  Labastida,
Marino and Vafaof \cite{LMV} have extended the results in \cite{OV} to
more non-trivial knots and links, and have described how to construct
Lagrangian submanifolds, for torus links at least, on the topological
string theory side of the duality.  It will be interesting to apply
our method to those cases as well.  Also, Aganagic and Vafa have
recently announced some interesting results on counting holomorphic
discs in Calabi--Yau 3--folds \cite{AV}, and we would like to
understand their results by means of relative stable morphisms.

\medskip
{\bf Acknowledgments}\qua The second author gratefully acknowledges
initial collaboration on this project and numerous useful discussions
with Jun S Song.  He also thanks Kefeng Liu for discussions and
correspondence.

The first author's research is supported in part by an NSF grant, a
Sloan fellowship and a Terman fellowship.  The second author's
research is supported in part by an NSF Graduate Fellowship and the
US Department of Energy under contract DE-AC03-76SF00515.

\bibliographystyle{gtart} \bibliography{link}

\end{document}